# Physical Layer Security: Coalitional Games for Distributed Cooperation

Walid Saad[1], Zhu Han[2], Tamer Başar[3], Mérouane Debbah[4], and Are Hjørungnes[1]
[1]UNIK - University Graduate Center, University of Oslo, Norway, {saad, arehj}@unik.no.
[2]Electrical and Computer Engineering Department, University of Houston, USA, zhan2@mail.uh.edu.
[3]Coordinated Science Laboratory, University of Illinois at Urbana-Champaign, USA, basar1@illinois.edu.
[4]Alcatel-Lucent Chair on Flexible Radio, SUPÉLEC, Gif-sur-Yvette, France, merouane.debbah@supelec.fr.

*Abstract*— **Cooperation between wireless network nodes is a promising technique for improving the physical layer security of wireless transmission, in terms of *secrecy capacity*, in the presence of multiple eavesdroppers. While existing physical layer security literature answered the question "what are the link-level secrecy capacity gains from cooperation?", this paper attempts to answer the question of "how to achieve those gains in a practical decentralized wireless network and in the presence of a secrecy capacity cost for information exchange?". For this purpose, we model the physical layer security cooperation problem as a coalitional game with non-transferable utility and propose a distributed algorithm for coalition formation. Through the proposed algorithm, the wireless users can autonomously cooperate and self-organize into disjoint independent coalitions, while maximizing their secrecy capacity taking into account the security costs during information exchange. We analyze the resulting coalitional structures, discuss their properties, and study how the users can self-adapt the network topology to environmental changes such as mobility. Simulation results show that the proposed algorithm allows the users to cooperate and self-organize while improving the average secrecy capacity per user up to $25.32\%$ relative to the non-cooperative case.**

## I. Introduction

During the past decade, security in wireless networks has been mainly considered at higher layers using various techniques such as cryptography. However, with the emergence of ad hoc and decentralized networks [1], [2], higher-layer techniques such as encryption are complex and hard to implement. Therefore, there has been a recent attention on studying the fundamental ability of the physical layer (PHY) to provide secure wireless communication. The main idea is to exploit the wireless channel PHY characteristics such as fading or noise for improving the reliability of wireless transmission. While these characteristics have always been seen as impairments, PHY layer security studies can utilize these characteristics for improving the security and reliability of wireless communication systems. This reliability is quantified by the *secrecy capacity*, which is defined as the maximum rate of secret information sent from a wireless node to its destination in the presence of eavesdroppers. The study of this security aspect began with the pioneering work of Wyner over the wire-tap channel [3] which showed that communicating data can be done in a secure manner without relying on any form of encryption such as secret keys. This work was followed up in [4], [5] for the scalar Gaussian wire-tap channel and the broadcast channel, respectively.

In recent years, there has been a growing interest in carrying out these studies unto the wireless and the multi-user channels [6–11]. For instance, in [6] and [7], the authors study the secrecy capacity region for both the Gaussian and the fading broadcast channels and propose optimal power allocation strategies. In [8], the secrecy level in multiple access channels from a link-level perspective is studied. Further, multiple antenna systems have been proposed in [9], [11] for ensuring a non-zero secrecy capacity, notably when the channel between the source and the destination is worse than the channel between the source and the eavesdroppers. Due to the size limitations of mobile devices, cooperation has been recently investigated as a practical way to achieve the multiple antenna gains [12]. In this context, the work in [10] investigates, through a two stage algorithm, the secrecy capacity gains (with no cost) resulting from the cooperation between a single cluster consisting of one source node and a number of relays. In this work, they investigate how a group of single antenna users can collaborate, by using beamforming (with no cost for cooperation), for improving their secrecy capacity. Briefly, the majority of the existing literature is devoted to the information theoretic analysis of link-level performance gains of secure communications with no information exchange cost, notably when a source node cooperate with some relays such as in [10]. No work seems to have investigated how a number of users, each with its own data, can interact and cooperate at network-wide level to improve their secrecy capacity and provide PHY security for their wireless transmission.

The main contribution of this work is to propose distributed cooperation strategies, through coalitional game theory, which allow to study the interactions between a network of users that seek to secure their communication through cooperation in the presence of multiple eavesdroppers. Another major contribution is to study the impact on the network topology and dynamics of the inherent trade off that exists between the PHY security cooperation gains in terms of secrecy capacity and the information exchange costs. In other words, while the earlier work answered the question "what are the secrecy capacity gains from cooperation?", here, we seek to answer the question of "how to achieve those gains in a practical decentralized wireless network and in the presence of a cost for information exchange?". We model

This work was done during the stay of Walid Saad at the Coordinated Science Laboratory, University of Illinois at Urbana-Champaign and was supported by the Research Council of Norway through projects 183311/S10, 176773/S10, and 18778/V11.

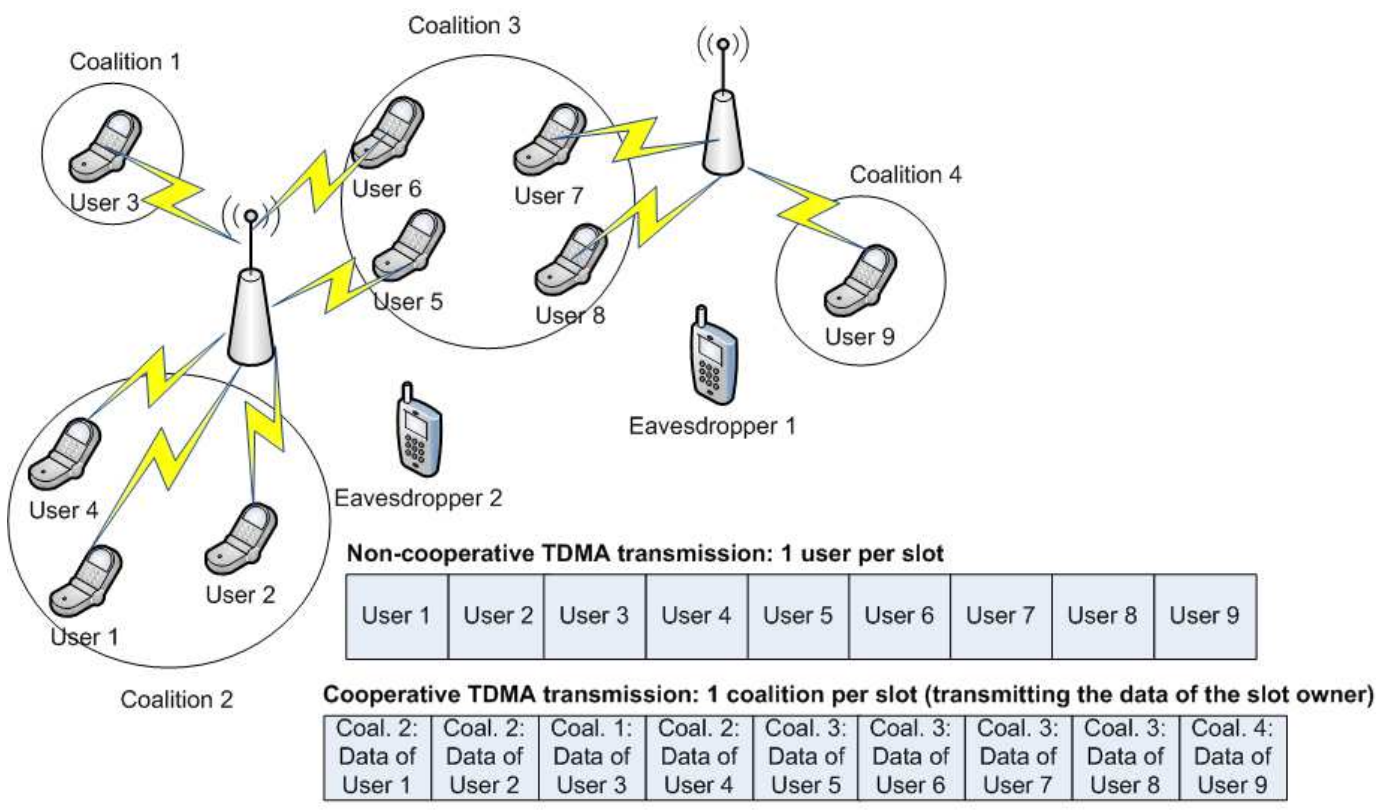

Fig. 1. System Model for Physical Layer Security Coalitional Game.

the problem as a non-transferable coalitional game and propose a distributed algorithm for autonomous coalition formation based on well suited concepts from cooperative games. Through the proposed algorithm, each user autonomously decides to form or break a coalition for maximizing its utility in terms of secrecy capacity while accounting for the loss of secrecy capacity during information exchange. We show that independent disjoint coalitions form in the network, due to the cooperation cost, and we study their properties. Through simulations, we assess the performance of the proposed algorithm, investigate the network topology, and show how the users can self-organize and adapt the topology to mobility. Simulation results show that the proposed algorithm allows the users to cooperate and self-organize while improving the average secrecy capacity per user up to $25.32\%$ relative to the non-cooperative case.

The rest of this paper is organized as follows: Section II presents the system model. Section III presents the game formulation and properties. In Section IV we devise the coalition formation algorithm. Simulation results are presented and analyzed in Section V. Finally, conclusions are drawn in Section VI.

## II. System Model

Consider a network having $N$ transmitters (e.g. mobile users) sending data to $M$ receivers (destinations) in the presence of $K$ eavesdroppers that seek to tap into the transmission of the users. Users, receivers and eavesdroppers are unidirectional-single-antenna nodes. We define $\mathcal{N} = \{1, \ldots, N\}$, $\mathcal{M} = \{1, \ldots, M\}$ and $\mathcal{K} = \{1, \ldots, K\}$ as the sets of users, destinations, and eavesdroppers, respectively. In this work, we consider only the case of multiple eavesdroppers, hence, we have $K > 1$. Furthermore, let $h_{i,m_i}$ denote the complex baseband channel gain between user $i \in \mathcal{N}$ and its destination $m_i \in \mathcal{M}$ and $g_{i,k}$ denote the channel gain between user $i \in \mathcal{N}$ and eavesdropper $k \in \mathcal{K}$. We consider a line of sight channel model with $h_{i,m_i} = d_{i,m_i}^{-\frac{\mu}{2}} e^{j\phi_{i,m_i}}$ with $d_{i,m_i}$ the distance between user $i$ and its destination $m_i$, $\mu$ the pathloss exponent, and $\phi_{i,m_i}$ the phase offset. A similar model is used for the user-eavesdropper channel.

For multiple access, we consider a TDMA transmission, whereby, in a non-cooperative manner, each user occupies a single time slot. Within a single slot, the maximum amount of reliable information transmitted from the user $i$ occupying the slot to its destination $m_i$ is quantified through the *secrecy capacity* $C_{i,m_i}$ defined as follows [6]:

$$C_{i,m_i} = \left( C_{i,m_i}^{d} - \max_{1 \leq k \leq K} C_{i,k}^{e} \right)^{+}, \qquad (1)$$

where $C_{i,m_i}^{d}$ is the Shannon capacity for the transmission between user $i$ and its destination $m_i \in \mathcal{M}$, $C_{i,k}^{e}$ is the Shannon capacity of user $i$ at the eavesdropper $k \in \mathcal{K}$, and $a^{+} \triangleq \max(a, 0)$.

In a non-cooperative approach, due to the broadcast nature of the wireless channel, the transmission of the users can be overheard by the eavesdroppers which reduces their secrecy capacity as clearly expressed in (1). For improving their performance and increasing their secrecy capacity, the users can collaborate by forming cooperative groups known as coalitions. Within every coalition, the users can utilize collaborative beamforming techniques for improving their secrecy capacities. In this context, every user $i$ belonging to a coalition $S$ will use the cooperation protocol of [10] by dividing its slot into two durations:

1) In the first duration, user $i$ broadcasts its information to the other members of coalition $S$.
2) In the second duration, coalitions $S$ performs collaborative beamforming. In other words, all the members of coalition $S$ relay a weighted version of user $i$'s signal to its destination.

The objective of this cooperation is to null the signal at the eavesdroppers, i.e., impose $C_{i,k}^{e} = 0, \forall k \in \mathcal{K}$, hence, improving the secrecy capacity in (1) [10]. Each coalition $S \subseteq \mathcal{N}$ that forms in the network is able to transmit within all the time slots previously held by its users. Thus, in the presence of cooperating coalitions, the TDMA system schedules one coalition per time slot. During a given slot, the coalition acts as a single entity for transmitting the data of the user that owns the slot. An illustration of this model is shown in Figure 1 for $N = 9$ users, $M = 2$ destinations, and $K = 2$ eavesdroppers.

Furthermore, we define a fixed transmit power *per time slot* $\tilde{P}$ which constrains *all the users* that are transmitting within a given slot. In a non-cooperative manner, this power constraint applies to the single user occupying the slot, while in a cooperative manner this *same* power constraint applies to the entire coalition occupying the slot. Such a power assumption is typical in TDMA systems comprising mobile users and is a direct result of ergodicity and the time varying user locations [12–14]. For every coalition $S$, during the time slot owned by user $i \in S$, user $i$ utilizes a portion of the available power $\tilde{P}$ for information exchange (first stage) while the remaining portion $P_i^S$ is used by the coalition $S$ to transmit the actual data to the destination $m_i$ of user $i$ (second stage). For information exchange, user $i \in S$ can broadcast its information to the farthest user $\hat{i} \in S$, by doing so all the other members of $S$ can also obtain the information due to the broadcast nature of the wireless channel. This information exchange incurs a power cost $\bar{P}_{i,\hat{i}}$ given by

$$\bar{P}_{i,\hat{i}} = \frac{\nu_0 \cdot \sigma^2}{q_{i,\hat{i}}}, \tag{2}$$

where $\nu_0$ is a target average signal-to-noise ratio (SNR) for information exchange, $\sigma^2$ is the noise variance and $q_{i,\hat{i}} = 1/d_{i,\hat{i}}^{\mu}$ is the path loss between users $i$ and $\hat{i}$ with $d_{i,\hat{i}}$ the distance between them. The remaining power that coalition $S$ utilizes for the transmission of the data of user $i$ during the remaining time of this user's slot is

$$P_i^S = (\tilde{P} - \bar{P}_{i,\hat{i}})^+ \tag{3}$$

For every coalition $S$, during the transmission of the data of user $i$ to its destination, the coalition members can weigh their signals in a way to *completely null* the signal at the eavesdroppers. We define, for a coalition $S$, the $|S| \times 1$ vectors $\boldsymbol{h}_S = [h_{i_1,m_1}, \ldots, h_{i_{|S|},m_{|S|}}]^H$, $\boldsymbol{g}_S^k = [g_{i_1,k}, \ldots, g_{i_{|S|},k}]^H$, and $\boldsymbol{w}_S = [w_{i_1}, \ldots, w_{i_{|S|}}]^H$ which represent, respectively, the "user-destination" channels, "user-eavesdropper $k$" channels, and the signal weights. By nulling the signals at the eavesdropper through cooperation within coalition $S$, the secrecy capacity (1) achieved by user $i \in S$ at its destination $m_i$ during user $i$'s time slot becomes [10, Eq. (14)]

$$C_{i,m_i}^S = \frac{1}{2}\log_2(1 + \frac{(\boldsymbol{w}_S^{opt})^H \boldsymbol{R}_S \boldsymbol{w}_S^{opt}}{\sigma^2}), \tag{4}$$

where $\boldsymbol{R}_S = \boldsymbol{h}_S \boldsymbol{h}_S^H$, $\sigma^2$ is the noise variance, and $\boldsymbol{w}_S^{opt}$ is the weight vector that maximizes the secrecy capacity while nulling the signal at the eavesdropper and is given in [10, Eq.(20)] by $\boldsymbol{w}_S^{opt} = \beta_i^S \boldsymbol{G}_S^H (\boldsymbol{G}_S \boldsymbol{G}_S^H)^{-1} \boldsymbol{e}$ with $\boldsymbol{G}_S = [\boldsymbol{h}_S, \boldsymbol{g}_S^1, \ldots, \boldsymbol{g}_S^K]^H$ a $(K+1) \times |S|$ matrix, $\beta_i^S = \sqrt{\frac{P_i^S}{\boldsymbol{e}^H (\boldsymbol{G}_S \boldsymbol{G}_S^H)^{-1} \boldsymbol{e}}}$ a scalar and $\boldsymbol{e} = [1, \boldsymbol{0}_{1\times K}]^H$ a $(K+1) \times 1$ vector. In (4), the factor $\frac{1}{2}$ accounts for the fact that half of the slot of user $i$ is reserved for information exchange.

Having adequately presented the model for physical layer security, the remainder of this paper is devoted to investigate how a network of users can cooperate, through the protocol described in this section, and improve the security of their wireless transmission, i.e., their secrecy capacity.

## III. Physical Layer Security as A Coalitional Game

In this section, we formulate the physical layer security model of the previous section as a coalitional game and we investigate its properties. For instance, the proposed PHY security problem is modeled as a coalitional game with a non-transferable utility defined as [15]:

**Definition 1:** A coalitional game with non-transferable utility is defined by a pair $(\mathcal{N}, V)$ where $\mathcal{N}$ is the set of players and $V$ is a mapping such that for every coalition $S \subseteq \mathcal{N}$, $V(S)$ is a closed convex subset of $\mathbb{R}^{|S|}$ that contains the payoff vectors that players in $S$ can achieve.

For the proposed physical layer security problem, given a coalition $S$ and denoting by $\phi_i(S)$ the payoff of user $i \in S$ during its time slot, we define the coalitional value set, i.e., the mapping $V$ as follows

$$V(S) = \{\boldsymbol{\phi}(S) \in \mathbb{R}^{|S|} | \ \forall i \in S \ \phi_i(S) = (v_i(S) - c_i(S))^+ \\ \text{if } P_i^S > 0, \text{ and } \phi_i(S) = -\infty \text{ otherwise.}\}, \tag{5}$$

where $v_i(S) = C_{i,m_i}^S$ is the gain in terms of secrecy capacity for user $i \in S$ given by (4) while taking into account the available power $P_i^S$ in (3) and $c_i(S)$ is a secrecy cost function that captures the loss for user $i \in S$, in terms of secrecy capacity, that occurs during information exchange. Note that, when *all* the power is spent for information exchange, the payoff $\phi_i(S)$ of user $i$ is set to $-\infty$ since, in this case, the user has clearly no interest in cooperating.

With regard to the secrecy cost function $c_i(S)$, when a user $i \in S$ sends its information to the farthest user $\hat{i} \in S$ using a power level $\bar{P}_{i,\hat{i}}$, the eavesdroppers can overhear the transmission. This security loss is quantified by the capacity at the eavesdroppers resulting from the information exchange and which, for a particular eavedropper $k \in \mathcal{K}$, is given by

$$\hat{C}_{i,k}^e = \frac{1}{2}\log(1 + \frac{\bar{P}_{i,\hat{i}} \cdot |g_{i,k}|^2}{\sigma^2}), \tag{6}$$

Given this security loss, the cost function $c(S)$ can be defined as

$$c_i(S) = \max(\hat{C}_{i,1}^e, \ldots, \hat{C}_{i,K}^e). \tag{7}$$

In a nutshell, the proposed coalitional value defined in (5) considers the benefit from cooperation, in terms of improved secrecy capacity, while taking into account the costs in terms of reduced power for transmission due to the power fraction used for information exchange as well as the secrecy capacity loss due to the eavesdroppers overhearing the transmission of the users during the information exchange phase.

Subsequently, the proposed physical layer security cooperation problem can be easily formulated as a coalitional game with non-transferable utility as per the following property:

**Property 1:** Given the mapping $V$ in (5), whenever the users transmit at their maximum rate (i.e. capacity), the proposed PHY security cooperation problem is a $(\mathcal{N}, V)$ non-transferable utility coalitional game.

*Proof:* Immediate result from the fact that when the users transmit at their maximum rate, the mapping $V(S)$ defined in (5) is a singleton set, and hence, closed and convex. Consequently, the proposed coalitional game model has a non-transferable utility $V(S)$ expressed by (5) ■

In general, coalitional game based problems seek to characterize the properties and stability of the grand coalition of all players since it is generally assumed that the grand coalition maximizes the utilities of the players [15]. In our case, although cooperation improves the secrecy capacity for the users in the TDMA network; the costs in terms of:

1) The fraction of power spent for information exchange as per (3) and,
2) the secrecy loss during information exchange as per (7)

strongly limit the cooperation gains. Therefore, for the proposed $(\mathcal{N}, v)$ coalitional game we have:

**Property 2:** For the proposed $(\mathcal{N}, V)$ coalitional game, the grand coalition of all the users *seldom* forms due to the various costs for information exchange. Instead, disjoint independent coalitions will form in the network.

*Proof:* Given a number of users positioned at different location within the wireless network, cooperation for improving the secrecy capacity entails costs, as previously mentioned, in terms of secrecy loss and power loss during information exchange as per (2) and (7). Hence, in a practical wireless network where the users are located at different positions, it is highly likely that, when they attempt to cooperate for forming the grand coalition $\mathcal{N}$ of all users, either: (i)- there exists a pair of users $i, j \in \mathcal{N}$ that are distant enough to require an information power cost of $\tilde{P}$ hence they have no incentive to join the grand coalition, or (ii)- there exists a user $i \in \mathcal{N}$ with the payoff of $i$ in the grand coalition $\phi_i(\mathcal{N}) = 0$ due to the secrecy loss as captured by (7), hence this user $i$ has incentive to deviate from the grand coalitions. Clearly, by accounting for the various cooperation costs, the grand coalition of all users will *seldom* form (it only forms if all users are very close, which is unrealistic in a large scale wireless network) and hence, the network structure consists of disjoint independent coalitions. ■

Due to this property, traditional solution concepts for coalitional games, such as the core [15], may not be applicable. In fact, in order for the core to exist, as a solution concept, a coalitional game must ensure that the grand coalition, i.e., the coalition of all players will form. However, as seen in Figure 1 and corroborated by Property 2, in general, due to the cost for coalition formation, the grand coalition will not form. Instead, independent and disjoint coalitions appear in the network as a result of the collaborative beamforming process. In this regard, the proposed game is classified as a *coalition formation game*, and the objective is to find the coalitional structure that will form in the network, instead of finding only a solution concept, such as the core, which aims mainly at stabilizing the grand coalition.

Furthermore, for the proposed $(\mathcal{N}, V)$ coalition formation game, a constraint on the coalition size, imposed by the nature of the cooperation protocol exists as follows:

**Remark 1:** For the proposed $(\mathcal{N}, V)$ coalition formation game, the size of any coalition $S \subseteq \mathcal{N}$ that will form in the network must satisfy $|S| > K$.

This is a direct result of the fact that, for nulling $K$ eavesdroppers, at least $K + 1$ users must cooperate, otherwise, no weight vector can be found to maximize (1) while nulling the signal at the eavesdroppers.

## IV. Distributed Coalition Formation Algorithm

### A. Coalition Formation Algorithm

Coalition formation has been a topic of high interest in game theory [16–19]. The goal is to find algorithms for characterizing the coalitional structures that form in a network where the grand coalition is not optimal. For instance, using game theoretical techniques from coalition formation games, we devise an algorithm for distributed coalition formation algorithm in the proposed $(\mathcal{N}, V)$ PHY security cooperative game. For constructing a coalition formation process suitable to the proposed game, we require the following definitions [18]

**Definition 2:** A *collection* of coalitions, denoted by $\mathcal{S}$, is defined as the set $\mathcal{S} = \{S_1, \ldots, S_l\}$ of mutually disjoint coalitions $S_i \subset \mathcal{N}$. In other words, a collection is any arbitrary group of disjoint coalitions $S_i$ of $\mathcal{N}$ not necessarily spanning all players of $\mathcal{N}$. If the collection spans *all* the players of $\mathcal{N}$; that is $\bigcup_{j=1}^{l} S_j = \mathcal{N}$, the collection is a *partition* of $\mathcal{N}$.

**Definition 3:** A preference operator or *comparison relation* $\rhd$ is an order defined for comparing two collections $\mathcal{R} = \{R_1, \ldots, R_l\}$ and $\mathcal{S} = \{S_1, \ldots, S_p\}$ that are partitions of the same subset $\mathcal{A} \subseteq \mathcal{N}$ (i.e. same players in $\mathcal{R}$ and $\mathcal{S}$). Therefore, $\mathcal{R} \rhd \mathcal{S}$ implies that the way $\mathcal{R}$ partitions $\mathcal{A}$ is preferred to the way $\mathcal{S}$ partitions $\mathcal{A}$.

Various well known orders can be used as comparison relations in different scenarios [18], [19]. These orders can be divided into two main categories: coalition value orders and individual value orders. Coalition value orders compare two collections (or partitions) using the value function of the coalitions inside these collections (suitable for games with transferable utilities) while individual value orders perform the comparison using the individual payoffs of every user. For the individual orders, two collections $\mathcal{R}$ and $\mathcal{S}$ are seen as two vectors of individual payoffs of the same length (corresponding to the total number of players) where each element of these payoff vectors corresponds to the utility received by the players in each coalition $R_i \in \mathcal{R}$ and $S_i \in \mathcal{S}$. In this context, individual value orders are quite suitable for non-transferable utility games such as the proposed game. Hence, for the PHY security coalition formation game, we define the following individual order that will be used in the coalition formation algorithm

**Definition 4:** Consider two collections $\mathcal{R} = \{R_1, \ldots, R_l\}$ and $\mathcal{S} = \{S_1, \ldots, S_m\}$ that are partitions of the same subset $\mathcal{A} \subseteq \mathcal{N}$ (same players in $\mathcal{R}$ and $\mathcal{S}$). For a collection $\mathcal{R} = \{R_1, \ldots, R_l\}$, let the utility of a player $j$ in a coalition $R_j \in \mathcal{R}$ be denoted by $\Phi_j(\mathcal{R}) = \phi_j(R_j) \in V(R_j)$. $\mathcal{R}$ is preferred over $\mathcal{S}$ by *Pareto order*, written as $\mathcal{R} \rhd \mathcal{S}$, iff

$$\mathcal{R} \rhd \mathcal{S} \Longleftrightarrow \{\Phi_j(\mathcal{R}) \geq \Phi_j(\mathcal{S}) \ \forall\ j \in \mathcal{R}, \mathcal{S}\},$$

with *at least one strict inequality* ($>$) for a player $k$.

In other words, a collection is preferred by the players over another collection, if at least one player is able to improve its payoff without hurting the other players. Subsequently, for performing autonomous coalition formation between the users in the proposed PHY security game, we construct a distributed algorithm based on two simple rules denoted as "merge" and "split" [18] defined as follows.

**Definition 5: Merge Rule -** Merge any set of coalitions $\{S_1, \ldots, S_l\}$ whenever the merged form is preferred by the players, i.e., where

$$\{\bigcup_{j=1}^{l} S_j\} \rhd \{S_1, \ldots, S_l\},$$

TABLE I

ONE ROUND OF THE PROPOSED PHY SECURITY COALITION FORMATION ALGORITHM

---

**Initial State**

The network is partitioned by $\mathcal{T} = \{T_1, \ldots, T_k\}$ (At the beginning of all time $\mathcal{T} = \mathcal{N} = \{1, \ldots, N\}$ with non-cooperative users).

**Three phases in each round of the coalition formation algorithm**

*Phase 1 - Neighbor Discovery:*

a) Each coalition surveys its neighborhood for candidate partners.
b) For every coalition $T_i$, the candidate partners lie in the area represented by the intersection of $|T_i|$ circles with centers $j \in T_i$ and radii determined by the distance where the power for information exchange does not exceed $\tilde{P}$ for any user (easily computed through (2)).

*Phase 2 - Adaptive Coalition Formation:*

In this phase, coalition formation using merge-and-split occurs.

**repeat**

a) $\mathcal{F}$ = Merge($\mathcal{T}$); coalitions in $\mathcal{T}$ decide to merge based on the algorithm of Section IV-A.
b) $\mathcal{T}$ = Split($\mathcal{F}$); coalitions in $\mathcal{F}$ decide to split based on the Pareto order.

**until** merge-and-split terminates.

*Phase 3 - Secure Transmission:*

Each coalition's users exchange their information and transmit their data within their allotted slots.

**The above three phases are repeated periodically during the network operation, allowing a topology that is adaptive to environmental changes such as mobility.**

---

therefore, $\{S_1, \ldots, S_l\} \to \{\bigcup_{j=1}^{l} S_j\}$.

**Definition 6: Split Rule -** Split any coalition $\bigcup_{j=1}^{l} S_j$ whenever a split form is preferred by the players, i.e., where

$$\{S_1, \ldots, S_l\} \triangleright \{\bigcup_{j=1}^{l} S_j\},$$

thus, $\{\bigcup_{j=1}^{l} S_j\} \to \{S_1, \ldots, S_l\}$.

Using the above rules, multiple coalitions can merge into a larger coalition if merging yields a preferred collection based on the Pareto order. This implies that a group of users can agree to form a larger coalition, if at least one of the users improves its payoff without decreasing the utilities of any of the other users. Similarly, an existing coalition can decide to split into smaller coalitions if splitting yields a preferred collection by Pareto order. The rationale behind these rules is that, once the users agree to sign a merge agreement, this agreement can only be broken if all the users approve. This is a family of coalition formation games known as coalition formation games with partially reversible agreements [16]. Using the rules of merge and split is highly suitable for the proposed PHY security game due to many reasons. For instance, each merge or split decision can be taken in a distributed manner by each individual user or by each already formed coalition. Further, it is shown in [18] that any arbitrary iteration of merge and split rules terminates, hence these rules can be used as building blocks in a coalition formation process for the PHY security game.

Accordingly, for the proposed PHY security game, we construct a coalition formation algorithm based on merge-and-split and divided into three phases: neighbor discovery, adaptive coalition formation, and transmission. In the neighbor discovery phase (Phase 1), each coalition (or user) surveys its environment in order to find possible cooperation candidates. For a coalition $S_k$ the area that is surveyed for discovery is the intersection of $|S_k|$ circles, centered at the coalition members with each circle's radius given by the maximum distance $\bar{r}_i$ (for the circle centered at $i \in S_k$) within which the power cost for user $i$ as given by (2) does not exceed the total available power $\tilde{P}$. This area is determined by the fact that, if a number of coalitions $\{S_1, \ldots, S_m\}$ attempt to merge into a new coalition $G = \cup_{i=1}^{m} S_i$ which contains a member $i \in G$ such that the power for information exchange needed by $i$ exceeds $\tilde{P}$, then the payoff of $i$ goes to $-\infty$ as per (5) and the Pareto order can never be verified. Clearly, as the number of users in a coalition increases, the number of circles increases, reducing the area where possible cooperation partners can be found. This implies that, as the size of a coalition grows, the possibility of adding new users decreases, and hence, the complexity of performing merge also decreases.

Following Phase 1, the adaptive coalition formation phase (Phase 2) begins, whereby the users interact for assessing whether to form new coalitions with their neighbors or whether to break their current coalition. For this purpose, an iteration of sequential merge-and-split rules occurs in the network, whereby each coalition decides to merge (or split) depending on the utility improvement that merging (or splitting) yields. Starting from an initial network partition $\mathcal{T} = \{T_1, \ldots, T_l\}$ of $\mathcal{N}$, any random coalition (individual user) can start with the merge process. The coalition $T_i \in \mathcal{T}$ which debuts the merge process starts by enumerating, sequentially, the possible coalitions, of size greater than $K$ (Remark 1), that it can form with the neighbors that were discovered in Phase 1. On one hand, if a new coalition $\tilde{T}_i$ which is preferred by the users through Pareto order is identified, this coalition will form by a merge agreement of all its members. Hence, the merge ends by a final merged coalition $T_i^{\text{final}}$ composed of $T_i$ and one or several of coalitions in its vicinity. On the other hand, if $T_i$ is unable to merge with any of the discovered partners, it ends its search and $T_i^{\text{final}} = T_i$.

The algorithm is repeated for the remaining $T_i \in \mathcal{T}$ until all the coalitions have made their merge decisions, resulting in a final partition $\mathcal{F}$. Following the merge process, the coalitions in the resulting partition $\mathcal{F}$ are next subject to split operations, if any is possible. In the proposed PHY security problem, the coalitions are only interested in splitting into structures that include either singleton users or coalitions of size larger than $K$ or both (Remark 1). Similar to merge, the split is a local decision to each coalition. An iteration consisting of multiple successive merge-and-split operations is repeated until it terminates. The termination of an iteration of merge and split rules is guaranteed as shown in [18]. It must be stressed that the merge or split decisions can be taken in a distributed way by the users/coalitions without relying on any centralized entity.

In the final transmission phase (Phase 3), the coalitions exchange their information and begin their secure transmission towards their corresponding destinations, in a TDMA manner, one coalition per slot. Every slot is owned by a user who transmits its data with the help of its coalition partners, if that user belongs to a coalition. Hence, in this phase, the user perform the actual beamforming, while transmitting the data of every user within its corresponding slot. Each run of the proposed algorithm consists of these three phases, and is summarized in Table I. As time evolves and the users, eavesdroppers and destinations move (or new users or eavesdroppers enter/leave the network), the users can autonomously self-organize and adapt the network's topology through appropriate merge-and-split decisions during Phase 2. This adaptation to environmental changes is ensured by enabling the users to run the adaptive coalition formation phase periodically in the network.

The proposed algorithm in Table I can be implemented in a distributed manner. As the user can detect the strength of other users' uplink signals (through techniques similar to those used in the ad hoc routing discovery), nearby coalitions can be discovered in Phase 1. In fact, during Phase 1, each coalition in the network can easily work out the area within which candidates for merge can be found, as previously explained in this section. Once the neighbors are discovered, the coalitions can perform merge operations based on the Pareto order in Phase 2. Moreover, each formed coalition can also internally decides to split if its members find a split form by Pareto order. By using a control channel, the distributed users can exchange some channel information and then and then cooperate using our model (exchange data information if needed, form coalition then transmit). Note that, in this paper, we assume that the users have perfect knowledge of the channels to the eavesdroppers which is an assumption used in most PHY security related literature, and as explained in [10] this channel information can be obtained by the users through a constant monitoring of the behavior of the eavesdroppers.

### *B. Partition Stability*

The result of the proposed algorithm in Table I is a network partition composed of disjoint independent coalitions. The stability of this network partition can be investigated using the concept of a defection function [18].

**Definition 7:** A *defection* function $\mathbb{D}$ is a function which associates with each partition $\mathcal{T}$ of $\mathcal{N}$ a group of collections in $\mathcal{N}$. A partition $\mathcal{T}=\{T_1,\ldots,T_l\}$ of $\mathcal{N}$ is $\mathbb{D}$*-stable* if no group of players is interested in leaving $\mathcal{T}$ when the players who leave can only form the collections allowed by $\mathbb{D}$.

We are interested in two defection functions [17–19]. First, the $\mathbb{D}_{hp}$ function which associates with each partition $\mathcal{T}$ of $\mathcal{N}$ the group of all partitions of $\mathcal{N}$ that can form through merge or split and the $\mathbb{D}_c$ function which associates with each partition $\mathcal{T}$ of $\mathcal{N}$ the group of all collections in $\mathcal{N}$. This function allows any group of players to leave the partition $\mathcal{T}$ of $\mathcal{N}$ through *any* operation and create an arbitrary *collection* in $\mathcal{N}$. Two forms of stability stem from these definitions: $\mathbb{D}_{hp}$ stability and a stronger $\mathbb{D}_c$ stability. A partition $\mathcal{T}$ is $\mathbb{D}_{hp}$-stable, if no player in $\mathcal{T}$ is interested in leaving $\mathcal{T}$ through merge-and-split to form other partitions in $\mathcal{N}$; while a partition $\mathcal{T}$ is $\mathbb{D}_c$-stable, if no player in $\mathcal{T}$ is interested in leaving $\mathcal{T}$ through *any* operation (not necessarily merge or split) to form other collections in $\mathcal{N}$.

Hence, a partition is $\mathbb{D}_{hp}$-stable if no coalition has an incentive to split or merge. For instance, a partition $\mathcal{T}=\{T_1,\ldots,T_l\}$ is $\mathbb{D}_{hp}$-stable, if the following two necessary and sufficient conditions are met [18], [17] ($\ntriangleright$ is the non-preference operator, opposite of $\triangleright$):

1) For each $i\in\{1,\ldots,m\}$ and for each partition $\{R_1,\ldots,R_m\}$ of $T_i\in\mathcal{T}$ we have
$$\{R_1,\ldots,R_m\}\ntriangleright T_i; \tag{8}$$
2) For each $S\subseteq\{1,\ldots,l\}$ we have
$$\bigcup_{i\in S} T_i \ntriangleright \{T_i|i\in S\}, \tag{9}$$

The above conditions are the generalized form (through the framework of [18]) of the $\mathbb{D}_{hp}$ stability conditions presented in [17]. Using this definition of $\mathbb{D}_{hp}$ stability, we have

**Theorem 1:** Every partition resulting from our proposed coalition formation algorithm is $\mathbb{D}_{hp}$-stable.

*Proof:* Consider a partition $\mathcal{T}$ resulting from the convergence of an iteration of merge-and-split operations such as in the algorithm of Table I, then no coalition in $\mathcal{T}$ can leave this partition through merge or split. For instance, assume $\mathcal{T}=\{T_1,\ldots,T_l\}$ is the partition resulting from the proposed merge-and-split algorithm. If for any $i\in\{1,\ldots,l\}$ and for any partition $\{S_1,\ldots,S_m\}$ of $T_i$ we assume that $\{S_1,\ldots,S_m\}\triangleright T_i$ then the partition $\mathcal{T}$ can still be modified through the application of the split rule on $T_i$ contradicting with the fact that $\mathcal{T}$ resulted from a termination of the merge-and-split iteration; therefore $\{S_1,\ldots,S_m\}\ntriangleright T_i$ (first $\mathbb{D}_{hp}$ stability condition verified). A similar reasoning is applicable in order to prove that $\mathcal{T}$ verifies the second condition; since otherwise a merge rule would still be applicable. ■

Furthermore, a $\mathbb{D}_c$-stable partition $\mathcal{T}$ is characterized by being a strongly stable partition, which satisfies the following properties:

1) A $\mathbb{D}_c$-stable partition is $\mathbb{D}_{hp}$-stable.
2) A $\mathbb{D}_c$-stable partition is a *unique* outcome of any iteration of merge-and-split.
3) A $\mathbb{D}_c$-stable partition $\mathcal{T}$ is a unique $\triangleright$-maximal partition, that is for all partitions $\mathcal{T}'\neq\mathcal{T}$ of $\mathcal{N}$, $\mathcal{T}\triangleright\mathcal{T}'$. In the case where $\triangleright$ represents the Pareto order, this implies that the $\mathbb{D}_c$-stable partition $\mathcal{T}$ is the partition that presents a *Pareto optimal* utility distribution for all the players.

Clearly, it is desirable that the network self-organizes unto a $\mathbb{D}_c$-stable partition. However, the existence of a $\mathbb{D}_c$-stable partition is not always guaranteed [18]. The $\mathbb{D}_c$-stable partition $\mathcal{T}=\{T_1,\ldots,T_l\}$ of the whole space $\mathcal{N}$ exists if a partition of $\mathcal{N}$ that verifies the following two necessary and sufficient conditions exists [18]:

1) For each $i\in\{1,\ldots,l\}$ and each pair of disjoint *coalitions* $S_1$ and $S_2$ such that $\{S_1\cup S_2\}\subseteq T_i$ we have $\{S_1\cup S_2\}\triangleright\{S_1,S_2\}$.

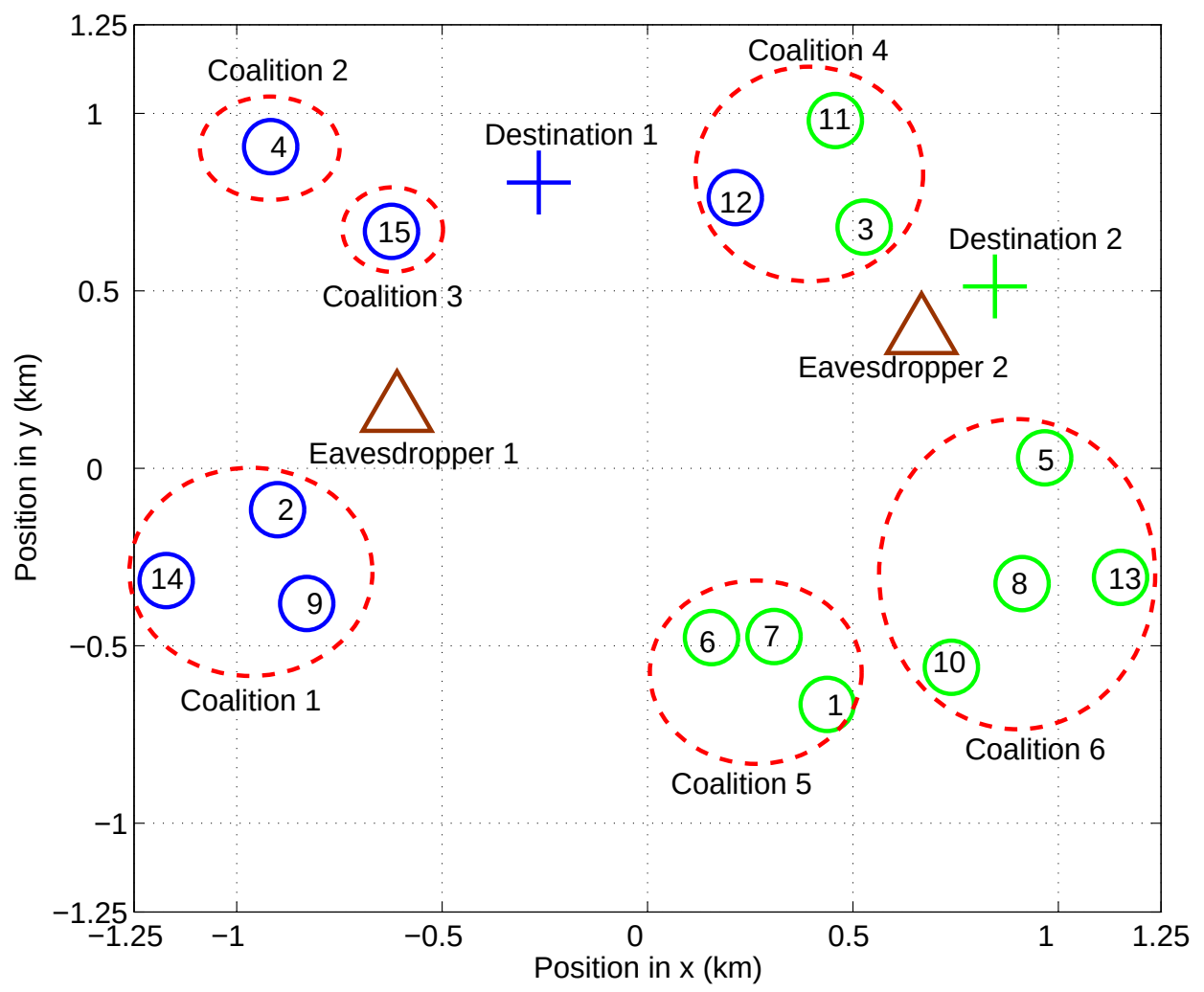

Fig. 2. A snapshot of a coalitional structure resulting from our proposed coalition formation algorithm for a network with $N = 15$ users, $M = 2$ destinations and $K = 2$ eavedroppers.

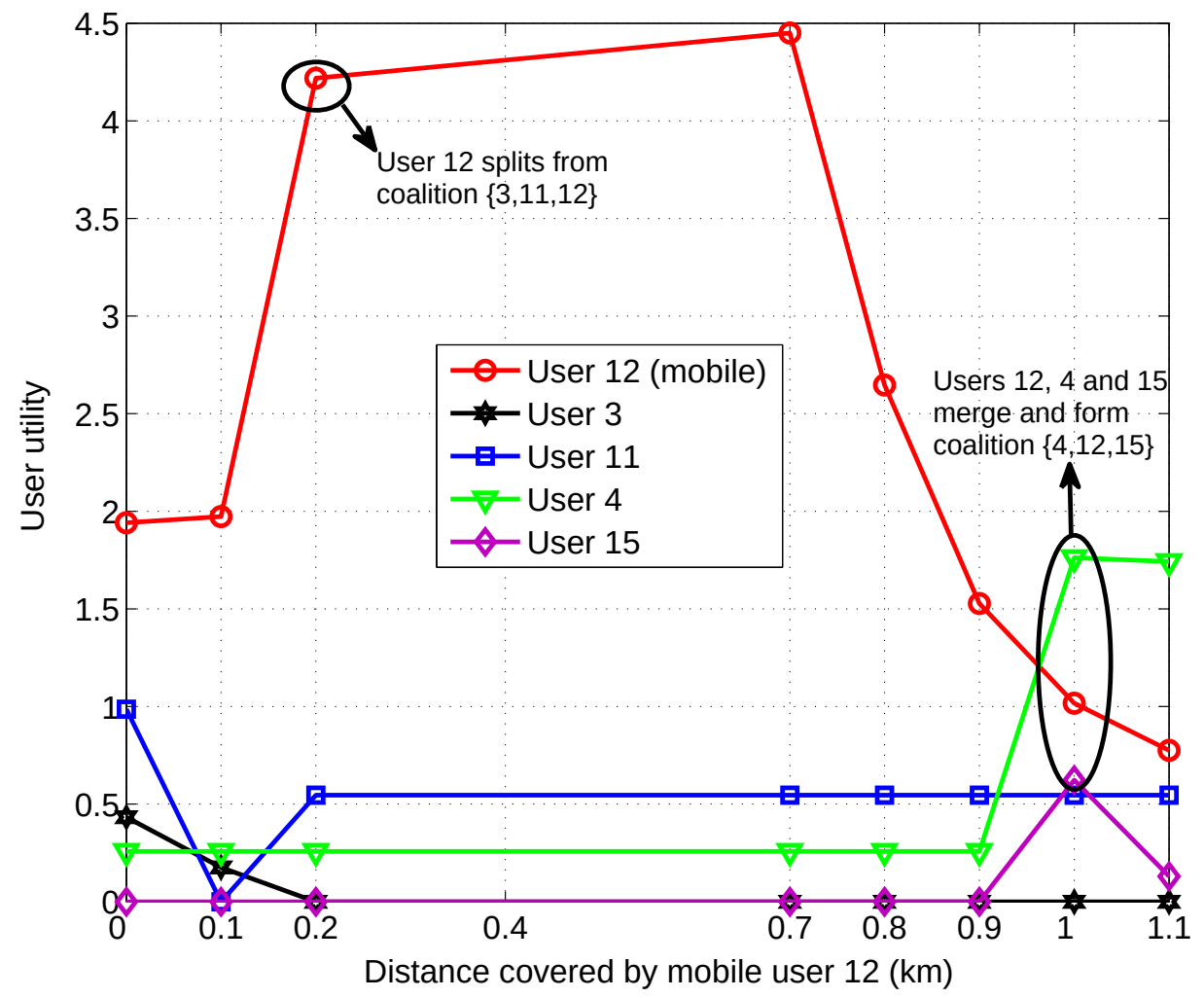

Fig. 3. Self-adaptation of the network's topology to mobility as User 12 in Figure 2 moves horizontally on the negative x-axis.

2) For the partition $\mathcal{T} = \{T_1, \ldots, T_l\}$ a coalition $G \subset \mathcal{N}$ formed of players belonging to different $T_i \in \mathcal{T}$ is $\mathcal{T}$-incompatible if for no $i \in \{1, \ldots, l\}$ we have $G \subset T_i$.

In summary, $\mathbb{D}_c$-stability requires that for all $\mathcal{T}$-incompatible coalitions $\{G\}[\mathcal{T}] \triangleright \{G\}$ where $\{G\}[\mathcal{T}] = \{G \cap T_i \ \forall\ i \in \{1, \ldots, l\}\}$ is the projection of coalition $G$ on $\mathcal{T}$. If no partition of $\mathcal{N}$ can satisfy these conditions, then no $\mathbb{D}_c$-stable partition of $\mathcal{N}$ exists. Nevertheless, we have

**Lemma 1:** For the proposed $(\mathcal{N}, v)$ PHY security coalitional game, the proposed algorithm of Table I converges to the optimal $\mathbb{D}_c$-stable partition, if such a partition exists. Otherwise, the final network partition is $\mathbb{D}_{hp}$-stable.

*Proof:* The proof is a consequence of Theorem 1 and the fact that the $\mathbb{D}_c$-stable partition is a unique outcome of any merge-and-split iteration [18] which is the case with any partition resulting from our algorithm. ■

Moreover, for the proposed game, the existence of the $\mathbb{D}_c$-stable partition cannot be always guaranteed. For instance, for verifying the first condition for existence of the $\mathbb{D}_c$-stable partition, the users that are members of each coalitions must verify the Pareto order through their utility given by (5). Similarly, for verifying the second condition of $\mathbb{D}_c$ stability, users belonging to all $\mathcal{T}$-incompatible coalitions in the network must verify the Pareto order. Consequently, the existence of such a $\mathbb{D}_c$-stable partition is strongly dependent on the location of the users and eavesdroppers through the individual utilities (secrecy capacities). Hence, the existence of the $\mathbb{D}_c$-stable partition is closely tied to the location of the users and the eavesdroppers, which, in a practical ad hoc wireless network are generally random. However, the proposed algorithm will always guarantee convergence to this optimal $\mathbb{D}_c$-stable partition when it exists as stated in Lemma 1. Whenever a $\mathbb{D}_c$-stable partition does not exist, the coalition structure resulting from the proposed algorithm will be $\mathbb{D}_{hp}$-stable (no coalition or individual user is able to merge or split any further).

## V. Simulation Results and Analysis

For simulations, a square network of $2.5$ km $\times$ $2.5$ km is set up with the users, eavesdroppers, and destinations randomly deployed within this area[1]. In this network, the users are always assigned to the closest destination, although other user-destination assignments can be used without any loss of generality in the proposed coalition formation algorithm. The simulation parameters used are as follows. First, the power constraint per slot is $\tilde{P} = 10$ mW, the noise level is $-90$ dBm, and the SNR for information exchange is $\nu_0 = 10$ dB which implies a neighbor discovery circle radius of $1$ km per user. For the channel model, the propagation loss is set to $\alpha = 3$.

In Figure 2, we show a snapshot of the network structure resulting from the proposed coalition formation algorithm for a randomly deployed network with $N = 15$ users, $M = 2$ destinations, and $K = 2$ eavesdroppers. This figure shows how the users self-organize into $6$ coalitions with the size of each coalition larger than $K$ or equal to $1$. For example, Users 4 and 15, having no suitable partners for forming a coalition of size larger than $2$, do not cooperate. The coalition formation process is a result of Pareto order agreements for merge (or split) between the users. For example, coalition $\{5, 8, 10, 13\}$ formed since all the users agree on its formation due to the fact that $V(\{5, 8, 10, 13\}) = \{\phi(\{5, 8, 10, 13\}) = [0.356\ \ 0.8952\ \ 1.7235\ \ 0.6213]\}$ which is a clear improvement on the non-cooperative utility which was $0$ for all four users (due to proximity to eavesdropper 2). In a nutshell, this figure shows how the users can self-organize into disjoint independent coalition for improving the PHY security of their wireless transmission.

In Figure 3 we show how the algorithm handles mobility through appropriate coalition formation decisions. For this purpose, the network setup of Figure 2 is considered while User

[1]This general network setting simulates a broad range of network types ranging from ad hoc networks, to sensor networks, WLAN networks as well as broadband or cellular networks.

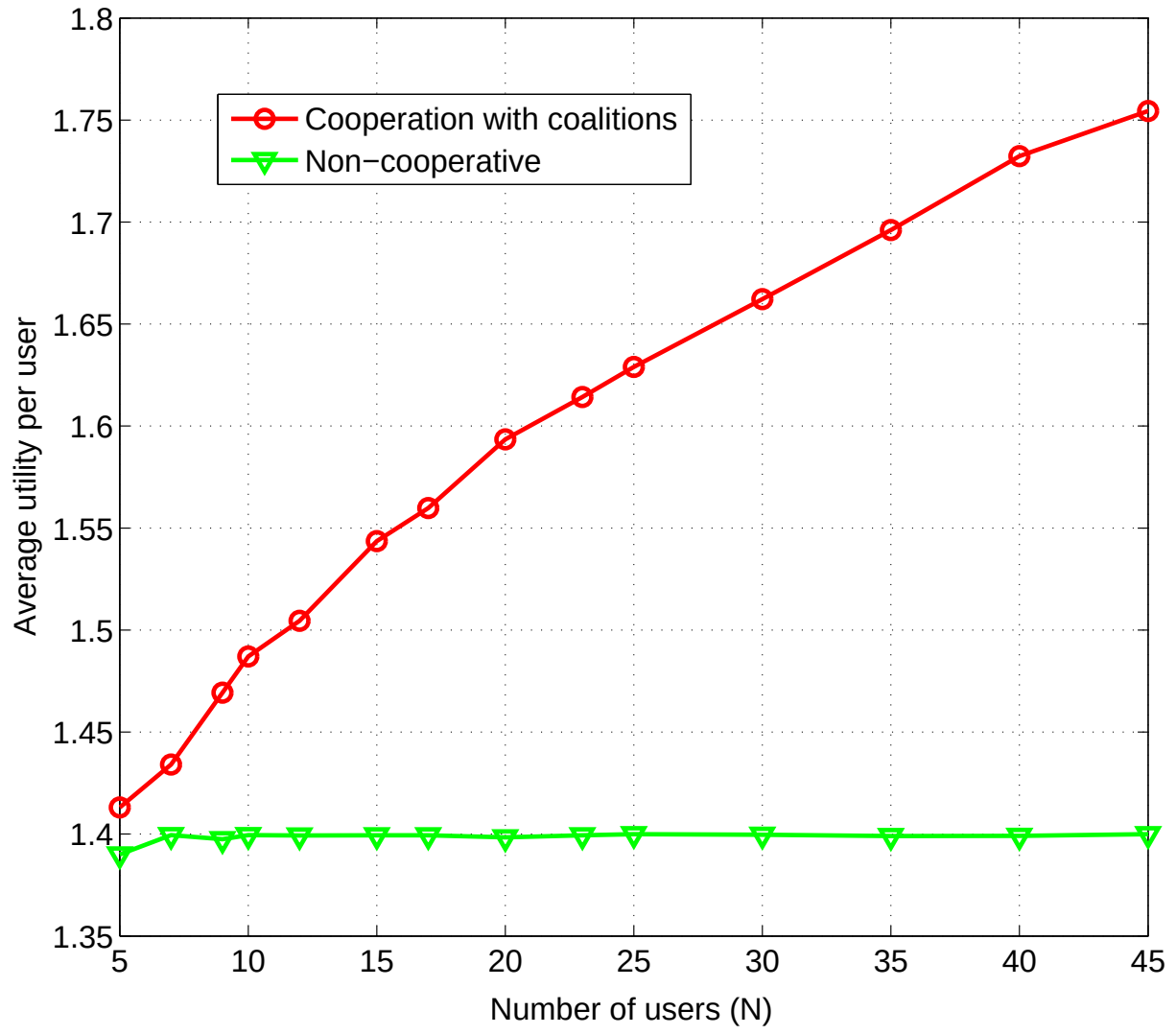

Fig. 4. Performance in terms of the average individual user utility (secrecy capacity) as a function of the network size $N$ for $M = 2$ destinations and $K = 2$ eavesdroppers.

12 is moving horizontally for 1.1 km in the direction of the *negative* x-axis. First of all, User 12 starts getting closer to its receiver (destination 2), and hence, it improves its utility. In the meantime, the utilities of User 12's partners (Users 3 and 11) drop due to the increasing cost. As long as the distance covered by User 12 is less than 0.2 km, the coalition of Users 1 and 6 can still bring mutual benefits to all three users. After that, splitting occurs by a mutual agreement and all three users transmit independently. When User 12 moves about 0.8 km, it begins to distance itself from its receiver and its utility begins to decrease. When the distance covered by User 12 reaches about 1 km, it will be beneficial to Users 12, 4, and 15 to form a 3-user coalition through the merge rule since they improve their utilities from $\phi_4(\{4\}) = 0.2577$, $\phi_{12}(\{12\}) = 0.7638$, and $\phi_{15}(\{15\}) = 0$ in a non-cooperative manner to $V(\{4, 12, 15\}) = \{\boldsymbol{\phi}(\{4, 12, 15\}) = [1.7618\ 1.0169\ 0.6227]\}$.

In Figure 4 we show the performance, in terms of average utility (secrecy capacity) per user, as a function of the network size $N$. The results are averaged over the random positions of the users, eavesdroppers and destinations. For cooperation with coalitions, the average individual utility increases with the number of users. This is interpreted by the fact that as the number of users $N$ increases, the probability of finding candidate partners to form coalitions with increases for every user. In contrast, for the non-cooperative approach an almost constant performance is noted. Finally, as easily seen in Figure 4 cooperation presents a clear performance advantage at all network sizes reaching up to $25.32\%$ improvement of the average user utility (secrecy capacity) at $N = 45$.

## VI. Conclusions

In this paper, we study the user behavior, topology, and dynamics of a network of users that interact in order to improve their secrecy capacity through cooperation. We formulate the problem as a non-transferable coalitional game, and propose a distributed and adaptive coalition formation algorithm. Through the proposed algorithm, the mobile users can autonomously take the decision to form or break cooperative coalitions through well suited rules from cooperative games while maximizing their secrecy capacity taking into account various costs for information exchange. We characterize the network structure resulting from the proposed algorithm, study its stability, and analyze the self-adaptation of the topology to environmental changes such as mobility. Simulation results show that the proposed algorithm allows the users to self-organize while improving the average secrecy capacity per user up to $25.32\%$ relative to the non-cooperative case.